\documentclass{osa-article}

\journal{osajournal}

\articletype{Research Article}

\begin{document}

\title{Can amplified spontaneous emission produce intense laser guide stars for adaptive optics?}

\author{Paul Hickson,\authormark{1,*} Joschua Hellemeier,\authormark{1} and Rui Yang\authormark{1,2}}

\address{\authormark{1}Department of Physics and Astronomy, University of British Columbia, 6224 Agriculture Road, Vancouver BC, V6T 1Z1,Canada\\
\authormark{2}School of Physics and Astronomy, Yunnan University, South Section, East Outer Ring Road, Chenggong District, Kunming, 650500, China
}

\email{\authormark{*}hickson@physics.ubc.ca} 



\begin{abstract}
Adaptive optics (AO) is a key technology for ground-based optical and infrared astronomy, providing high angular resolution and sensitivity. AO systems employing laser guide stars (LGS) can achieve high sky coverage, but their performance is limited by LGS return flux. We examine the potential of two new approaches that might produce high-intensity atmospheric laser beacons. Amplified spontaneous emission could potentially boost the intensity of beacons produced by conventional resonant excitation of atomic or molecular species in the upper atmosphere. This requires the production of a population inversion in an electronic transition that is optically-thick to stimulated emission. Potential excitation mechanisms include continuous wave pumping, pulsed excitation and plasma generation. Alternatively, a high-power femtosecond pulsed laser could produce a white-light supercontinuum high in the atmosphere. The broad-band emission from such a source could also facilitate the sensing of the tilt component of atmospheric turbulence.
\end{abstract}

\section{Introduction}

Adaptive optics (AO) is an important technology employed by many ground-based telescopes to compensate atmospheric turbulence, thereby greatly improving the resolution of astronomical images \cite{Babcock1953}. Many AO systems employ laser guide star (LGS) beacons that allow sensing of the turbulence in any desired direction, increasing sky coverage \cite{Foy1985}. Multiple beacons can be used to improve performance and increase the size of the corrected field of view \cite{Beckers1989,Murphy1991,Ellerbroek1994,Neichel2014}.  Large telescopes typically employ resonant LGS systems in which mesospheric atoms are excited, typically the $D_2$ transition of neutral  sodium \cite{Thompson1987}.  

One factor limiting the performance of such AO systems is the low brightness of the LGS. Much research has been devoted to optimizing the return flux of sodium LGS \cite{Holzlohner2010,Rampy2015}, and it appears that the limit of this technology is being approached. This motivates us to look for alternative ways to create LGS.

One interesting possibility is to induce amplified spontaneous emission (ASE) \cite{Milonni2010} in order to produce very bright LGS. Laser pumping of an abundant atmospheric atomic or molecular species could in principle produce a population inversion that would amplify radiation generated by spontaneous emission, as it propagates back towards the telescope. ASE in sodium vapour has been demonstrated in laboratory experiments \cite{Bustos2018}.

A second possible new approach is to use a high-power femtosecond pulsed laser to generate white-light supercontinuum (SC) emission high in the atmosphere \cite{Rairoux2000,Kasparian2000}. This might help address a limitation of conventional monochromatic laser AO in which the tilt component of atmospheric turbulence cannot be sensed because the deflection of downward propagating light from the LGS is equal and opposite to the deflection of the upward propagating radiation that creates the beacon. This degeneracy is normally broken by observations of stars in the field of view, but such stars are generally very faint, limiting the accuracy that can be achieved. The creation of a polychromatic LGS, which emits light at two or more widely-separated wavelengths, has been proposed to solve this problem \cite{Foy1995,Pique2006,deChatellus2008}, but has not been demonstrated in practice. 

In this paper we examine the physical processes of ASE and SC with the aim of assessing the feasibility of applying these effects in astronomical adaptive optics, and determining the conditions that would be required to produce useful AO beacons.

\section{Amplified spontaneous emission}

The essential physics of ASE can be captured by consideration of an atomic or molecular species that has a two-level transition, placed in a radiation field of specific energy density $\rho_\nu(\nu)$. Let $\nu_0$ be the central frequency of the transition. If non-radiative excitation and de-excitation can be ignored, the average fraction $x$ of atoms in the upper states is described by the rate equation
\begin{align}
  \frac{dx}{dt} & = -xA_{21} - \rho_\nu[x B_{21} - (1-x)B_{12}] \nonumber \\
  & = -A_{21} \left\{ x +\frac{\lambda^3\rho_\nu}{8\pi h}\left[x-\frac{g_2}{g_1}(1-x)\right]\right\}, \label{eq:rate}
\end{align}
where $A_{21}$, $B_{21}$ and $B_{12}$ are the Einstein coefficients for spontaneous and induced emission and absorption \cite{Milonni2010}. In obtaining the second line of Eq (\ref{eq:rate}), we have used the usual relations between the Einstein coefficient, $g_1B_{12} = g_2B_{21}$ and $ B_{21} = A_{21}\lambda^3/8\pi h$, where $\lambda = c/\nu$ is the wavelength and $g_1$ and $g_2$ are the statistical weights of the lower and upper states respectively.

In the limit of high intensity illumination, the excitation fraction reaches an equilibrium value $x_e$, that can be found by equating the LHS of Eq. (\ref{eq:rate}) to zero,
\begin{equation}
  x_e = g_2/(g_1+g_2).
\end{equation}

The specific intensity $I_\nu$ of radiation propagating through the medium is described by the equation of radiative transfer \cite{Chandrasekhar1950},
\begin{equation}
  \frac{dI_\nu}{ds} = -\kappa I_\nu + j_\nu. \label{eq:transfer}
\end{equation}
Here $\kappa$ is the absorption coefficient, $j_\nu$ is the emission coefficient, and $s$ is distance measured along the propagation path. These coefficients are related to the atomic number density $n$, excitation fraction $x$, and the Einstein coefficients by
\begin{align}
  j_\nu &= \frac{h\nu}{4\pi} A_{21}xn\varphi, \label{eq:j} \\
  \kappa & = \frac{h\nu}{c} [(1-x)B_{12}-xB_{21}]n\varphi = \frac{\lambda^2}{8\pi} y A_{21}n\varphi. \label{eq:kappa} 
\end{align}
Here $\varphi(\nu)$ is the line profile, normalized to have unit integral, and
\begin{equation}
  y = \frac{x_e-x}{1-x_e} = \frac{g_2}{g_1}(1-x)-x. \label{eq:y}
\end{equation}

If the excitation fraction can be increased above the equilibrium value, ($x > x_e$), producing a population inversion, the absorption coefficient becomes negative and the intensity of propagating radiation grows until the radiation escapes from the medium. If the excited region is a long narrow cylinder, as for excitation by a pump laser launched from a telescope, forward and backward propagating beams would develop. The intensity of these beams could in principle be very large, ultimately limited only by the power absorbed by the medium from the pump laser.

\section{LGS return flux}

The flux of radiation returning to the telescope depends on the atomic parameters, column density, and the nature of the excitation. For emission generated within the medium, Eq (\ref{eq:transfer}) has the solution
\begin{equation}
  I_\nu(\tau) = e^{-\tau}\int_0^\tau S_\nu(u) e^udu, \label{eq:I}
\end{equation}
where
\begin{equation}
  \tau = \int \kappa ds,
\end{equation}
is the optical depth and
\begin{equation}
  S_\nu = \frac{j_\nu}{\kappa} = -\frac{2h\nu^3}{c^2}\frac{x}{y}.
\end{equation}
is the source function. If the excitation fraction $x$ is constant within the emitting region, $S_\nu$ is constant and can be taken outside the integral, which is then easily evaluated. The result is
\begin{equation}
  I_\nu = \frac{2h\nu^3}{c^2}\frac{x}{y}(1-e^{-\tau}). \label{eq:I_nu}
\end{equation}
The optical depth becomes
\begin{equation}
    \tau =  \frac{\lambda^2}{8\pi}y  NA_{21} \varphi, \label{eq:tau}
\end{equation}
where 
\begin{equation}
  N = \int_0^\infty n ds
\end{equation}
is the column density of the atomic or molecular species.

Photons arriving at the telescope will be confined to a solid angle $\Omega \simeq \mathcal{A}/z^2$, where $\mathcal{A}$ is the transverse area of the emitting region and $z$ is the line-of-sight distance to its center. The photon flux (photons s$^{-1}$ m$^{-2}$) is found by dividing the intensity by the photon energy $h\nu$ and integrating over solid angle and frequency. Thus,
\begin{equation}
  \Phi \simeq \frac{\Omega}{h\nu} \int_0^\infty I_\nu d\nu \simeq \frac{\mathcal{A}}{h\nu z^2} \int_0^\infty I_\nu d\nu. \label{eq:Phi}
\end{equation}

Substituting Eqs (\ref{eq:I_nu}) and (\ref{eq:tau}) into Eq. (\ref{eq:Phi}) and setting $x = 1$, we obtain
\begin{equation}
  \Phi = -\frac{2\Omega}{c^2}\frac{x}{y}\int_0^\infty \left\{\exp\left[ -\frac{c^2}{8\pi\nu^2} yNA_{21}\varphi(\nu)  \right]-1\right\} \nu^2 d\nu. \label{eq:Phi3}
\end{equation}
The line profile is a symmetric function that is sharply peaked about the transition frequency $\nu_0$, so little error is introduced by replacing $\nu$ by $\nu_0$ in the coefficient of $\varphi$. We can also change the variable of integration to $q = \nu-\nu_0$ and make use of use the symmetry to obtain
\begin{equation}
  \Phi = -\frac{4\Omega}{\lambda_0^2}\frac{x}{y}\int_0^\infty \left[e^{-\tau_0\zeta} -1\right] dq. \label{eq:Phi4}
\end{equation}
where $\zeta(q) = \varphi(q)/\varphi_0$, $\varphi_0 \equiv \varphi(\nu_0)$ and 
\begin{equation}
\tau_0 = \frac{\lambda_0^2}{8\pi}yNA_{21}\varphi_0 \label{eq:tau_0}
\end{equation}
is the optical depth at the line center. 

We start by expanding the exponential and integrating term by term,
\begin{equation}
  \Phi = -\frac{4\Omega}{\lambda_0^2}\frac{x}{y}\sum_{k=1}^\infty \frac{(-\tau_0)^k}{k!}\int_0^\infty \zeta^k dq. \label{eq:Phi4}
\end{equation}
For a Lorentzian line profile, of frequency half-width $w$ (half-width at half maximum intensity),
\begin{align}
  \zeta_L & = [1+(q/w)^2]^{-1},  \label{eq:zeta_L}\\
  \varphi_0 & = 1/(\pi w). \label{eq:varphi_0}
\end{align}
Substituting Eq. (\ref{eq:zeta_L}) into Eq. (\ref{eq:Phi4}), we obtain
\begin{equation}
  \Phi = -\frac{4\Omega}{\lambda_0^2}\frac{x}{y}\sum_{k=1}^\infty \frac{(-\tau_0)^k}{k!}\int_0^\infty [1+(q/w)^2]^{-k} dq. \label{eq:Phi_L}
\end{equation}
The integral is a Mellin transform and can be expressed in terms of gamma functions $\Gamma(x)$, 
\begin{equation}
  \Phi = -\frac{2\Omega}{\lambda^2}\frac{x}{y}\sum_{k=1}^\infty \frac{(-\tau_0)^k}{k!} \frac{\Gamma(1/2)\Gamma(k-1/2)}{\Gamma(k)}.
\end{equation}
Recalling that $\Gamma(1/2) = \sqrt{\pi}$ and $k! = \Gamma(k+1)$, we can express this in terms of a generalized hypergeometric function $_pF_q$. Making the substitution $k \to k+1$ we find 
\begin{align}
  \Phi & =  \frac{2\sqrt{\pi}\tau_0w\Omega}{\lambda^2}\frac{x}{y}\sum_{k=0}^\infty  \frac{\Gamma(k+1/2)}{\Gamma(k+2)} \frac{(-\tau_0)^k}{k!} \nonumber \\
   & = \frac{2\pi\tau_0w\Omega}{\lambda^2}\frac{x}{y} ~_1F_1(1/2;2;-\tau_0).
\end{align}
Substituting from Eqs. (\ref{eq:tau_0}) and (\ref{eq:varphi_0}) and simplifying, we obtain
\begin{equation}
    \Phi = -\frac{\mathcal{A}}{4\pi z^2} xNA_{21} ~_1F_1(1/2;2;-\tau_0), \label{Phi6}
\end{equation}
This extends previous work by Milonni and Eberly, who give an approximate solution \cite{Milonni2010}. 

For the case of complete excitation (all atoms in the upper state), $y = -1$, and there is a near-exponential dependence of the flux on the magnitude of the optical depth.  The first few terms of the hypergeometric series are, (setting $y = -1$), 
\begin{equation}
    \Phi = \frac{xN\mathcal{A}A_{21}}{4\pi z^2}  \left[ 1 - \frac{1}{8} \tau_0 + \frac{1}{48} \tau_0^2 + \cdots  \right]. \\
\end{equation}
The numerator of the fraction is the total number of spontaneous transitions per second from all excited atoms. The first term in the series can thus be recognized as the flux from spontaneous emission alone, and subsequent terms generate the exponential growth with optical depth. It is clear from this that \emph{if the transition is optically-thin ($|\tau_0| \ll 1$) there is essentially no ASE gain}. 

It is convenient to express the optical depth in terms of the cross section $\sigma_{21}$ for stimulated emission, defined by
\begin{equation}
  \sigma_{21}(\nu) = \frac{h\nu}{c}B_{21}\varphi(\nu) = \frac{\lambda^2}{8\pi}A_{21}\varphi(\nu). \label{eq:sigma_21}
\end{equation}
With this definition, the optical depth at the line centre becomes
\begin{equation}
  \tau_0 =  y N\sigma_{21}(\nu_0), \label{eq:tau2} \\
\end{equation}
We see that \emph{the critical parameters that determine the optical depth are the excitation fraction, column density and transition cross section}. 

A similar analysis can be done for a Gaussian line profile, 
\begin{align}
  \zeta & = \exp[-\ln 2\, (q/w)^2], \\
  \varphi_0 & = \sqrt{\ln 2/(\pi w^2)}
\end{align}
In that case, the flux is given by
\begin{align}
  \Phi & = \frac{\Omega}{4\pi}xNA_{21}\xi(-\tau_0), \label{eq:Phi_G} \\
  & \simeq \frac{xN\mathcal{A}A_{21}}{4\pi z^2}  \left[ 1 - \frac{1}{2\sqrt{2}} \tau_0 + \frac{1}{6\sqrt{3}} \tau_0^2 + \cdots  \right],
\end{align}
The function
\begin{equation}
  \xi(t) \equiv \sum_{k=0}^\infty \frac{t^k}{(k+1)^{1/2}(k+1)!}
\end{equation}
approaches an exponential as $t \to \infty$.

\section{Excitation mechanisms}

A prerequisite to achieving ASE is a mechanism that generates a population inversion in the medium. 

\subsection{CW excitation}

A possible method of excitation is to use a continuous wave (CW) laser to pump atoms to the upper state. In order to produce a population inversion, one needs at least three transitions that include a metastable level. This level is populated by spontaneous transitions from a higher level that is pumped by the laser. A difficulty with this approach is that the metastable level will necessarily have a comparatively-low cross section, as shown by Eq. (\ref{eq:sigma_21}). Thus, the optical depth will be small, unless the column density is very large. 

A second problem is the relatively-narrow linewidth of CW lasers, typically on the order of $10^7$ Hz. This is roughly 1\% of the Doppler width of $\sim 1$ GHz, which means that only about 1\% of the available atoms can be excited. The effective column density, and therefore the optical depth, is reduced by this same factor. 

\subsection{Coherent excitation}

A second possible approach is coherent excitation in which a pulsed laser is used to induce Rabi oscillations in the atomic transition. For a $\pi$-pulse, the product of the laser irradiance $F$ and pulse length $\tau_p$ is such that one half of a Rabi oscillation occurs, inverting the population. Atoms that are in the lower state are transferred to the upper state and vice versa. The condition for this is \cite{Milonni1992}
\begin{equation}
  F\tau_p^2 = \hbar c\pi^4/(3\lambda_0^3A_{21}).
\end{equation}
For an optical transition of a typical strong resonance line, and a 100 ps pulse length, an irradiance on the order of $10^7$ W $m^{-2}$ is required. For a beam area of 1 m$^{2}$, this corresponds to a pulse energy of 1 mJ, which is readily achievable.

A pulsed laser delivering a train of $\pi$-pulses could repeatedly excite the atomic population, creating a population inversion. However, to support ASE, this population inversion must be maintained for a time comparable to the propagation time through the mesosphere, on the order of $10^{-4}$ s. This puts an upper limit on the transition rate of $A_{21} \lesssim 10^4$. That results in a very-low optical depth for all mesospheric metals \cite{Yang2020}

\subsection{High-power pulsed lasers}

A third approach is offered by high-power femtosecond pulsed lasers. Laser pulses can induce ionization of atmospheric atoms and molecules if the electric field in the pulse is comparable to the Coulomb field of the nucleus at the atomic radius \cite{Corcum1988,Chin1988}. This occurs at an irradiance of $F \sim 10^{18}$ W m$^{-2}$. If the laser pulse is pre-chirped (time-dependent frequency) it can be intensified by atmospheric dispersion as it propagates, producing a plasma at a chosen distance \cite{Rairoux2000}.  The desired excitation state might then be achieved during recombination.

\section{Supercontinuum emission}

The advent of terrawatt femtosecond lasers offers intriguing new possibilities for adaptive optics. A ubiquitous feature of the interaction of very intense pulses with continuous media, including gases, is the generation of white-light ``supercontinuum'' (SC) emission \cite{Alfano1970a,Alfano1970b}. The exact nature of this radiation is not well understood, but likely arises from a combination of nonlinear effects \cite{Alfano2005}. In the atmosphere, filaments of emission have been generated that can extend for several km \cite{Woste1997,Corcum1989,Rairoux2000}. Emission from these filaments is found to be highly directional, peaking in the forward and backscattered direction. The intensity of the backward-propagating beam is found to be enhanced compared to the predicted linear Rayleigh-Mie scattering theory. This is presumed to be a result of non-linear reflection arising from longitudinal index-of-refraction variations produced by Kerr and plasma effects \cite{Yu2001}. 

Remote SC filaments can be generated by chirping the transmitted pulses. Simulations indicate that pulses of sufficient intensity and coherence can be formed at altitudes as high as 20 km \cite{Kasparian2000,Kasparian2003,Theberge2008}.

The white-light emission observed in SC radiation is believed to result from self-phase modulation (SPM) due to nonlinear interaction with the medium. SPM occurs when the pulse induces a change in the index of refraction in the medium due to the optical Kerr effect. This induces a chirp in the pulse, broadening its frequency spectrum. The degree of spectral broadening increases with increasing laser power. Spectra of atmospheric SC filaments show emission that extends over the entire visible range from the ultraviolet to the infrared \cite{Rairoux2000,Kasparian2003}. 

\section{Discussion and conclusions}

We have estimated the photon flux that could in principle be provided by ASE, if a suitable atomic or molecular population, transition, and excitation mechanism can be found. We conclude that no significant ASE gain can arise if the transition is optically-thin. For ASE, the magnitude of the optical depth must exceed unity, ideally by an order of magnitude. The optical depth depends only on the excitation fraction, the column density and the transition cross section for stimulated emission. The metallic species present in the mesosphere do not have sufficient column density to support ASE \cite{Yang2020}.


High-power pulsed lasers that can create a plasma filament and superradiance in the atmosphere may provide an opportunity to realize high-intensity white-light laser guide stars. If feasible, this would allow sensing of atmospheric tilt, in addition to higher-order wavefront distortion. A potential drawback is the relatively low altitude of the filaments that have so far been produced, which increases the ``cone effect''. However, this can be mitigated by the use of multiple laser guide stars. 

\section*{Acknowledgements}

PH acknowledges financial support from the Natural Sciences and Engineering Research Council of Canada. He thanks NAOC for hospitality during a sabbatical visit made possible by financial support from the Chinese Academy of Sciences, via the CAS Presidents International Fellowship Initiative, 2017VMA0013. RY acknowledges financial support by Yunnan Provincial Department of Education for her visiting scholar position at UBC.


\end{document}